\documentclass[3p,preprint,12pt]{elsarticle}
\makeatletter\if@twocolumn\PassOptionsToPackage{switch}{lineno}\else\fi\makeatother

\usepackage{tabulary,xcolor}
\usepackage{amsfonts,amsmath,amssymb}
\usepackage[T1]{fontenc}
\makeatletter
\let\save@ps@pprintTitle\ps@pprintTitle
\def\hlinewd#1{%
  \noalign{\ifnum0=`}\fi\hrule \@height #1%
  \futurelet\reserved@a\@xhline}

\AtBeginDocument{\ifNAT@numbers \biboptions{sort&compress}\fi}
\makeatother

\usepackage{natbib}
\setcitestyle{authoryear,open={(},close={)}} 

\usepackage{bbm}

\usepackage{courier}

\usepackage{hyperref}
\usepackage{subcaption}

\usepackage{ifluatex}
\ifluatex
\usepackage{fontspec}
\defaultfontfeatures{Ligatures=TeX}
\usepackage[]{unicode-math}
\unimathsetup{math-style=TeX}
\else 
\usepackage[utf8]{inputenc}
\fi 
\ifluatex\else\usepackage{stmaryrd}\fi

\usepackage{url,multirow,morefloats,floatflt,cancel,tfrupee}
\makeatletter

\AtBeginDocument{\@ifpackageloaded{textcomp}{}{\usepackage{textcomp}}}
\makeatother
\usepackage{colortbl}
\usepackage{xcolor}
\usepackage{pifont}
\usepackage[nointegrals]{wasysym}
\makeatletter

\def\mcWidth#1{\csname TY@F#1\endcsname+\tabcolsep}

\def\cAlignHack{\rightskip\@flushglue\leftskip\@flushglue\parindent\z@\parfillskip\z@skip}
\def\rAlignHack{\rightskip\z@skip\leftskip\@flushglue \parindent\z@\parfillskip\z@skip}

\@ifundefined{etal}{}{}

\usepackage{ifxetex}
\ifxetex\else\if@twocolumn\@ifpackageloaded{stfloats}{}{\usepackage{dblfloatfix}}\fi\fi

\AtBeginDocument{
\expandafter\ifx\csname eqalign\endcsname\relax
\def\eqalign#1{\null\vcenter{\def\\{\cr}\openup\jot\m@th
  \ialign{\strut$\displaystyle{##}$\hfil&$\displaystyle{{}##}$\hfil
      \crcr#1\crcr}}\,}
\fi
}

\AtBeginDocument{%
  \@ifpackageloaded{endfloat}%
   {\renewcommand\efloat@iwrite[1]{\immediate\expandafter\protected@write\csname efloat@post#1\endcsname{}}}{\newif\ifefloat@tables}%
}%

\def\BreakURLText#1{\@tfor\brk@tempa:=#1\do{\brk@tempa\hskip0pt}}
\let\lt=<
\let\gt=>
\def\processVert{\ifmmode|\else\textbar\fi}

\@ifundefined{subparagraph}{
\def\subparagraph{\@startsection{paragraph}{5}{2\parindent}{0ex plus 0.1ex minus 0.1ex}%
{0ex}{\normalfont\small\itshape}}%
}{}

\newcommand\role[1]{\unskip}
\newcommand\aucollab[1]{\unskip}
  
\@ifundefined{tsGraphicsScaleX}{\gdef\tsGraphicsScaleX{1}}{}
\@ifundefined{tsGraphicsScaleY}{\gdef\tsGraphicsScaleY{.9}}{}
\def\checkGraphicsWidth{\ifdim\Gin@nat@width>\linewidth
	\tsGraphicsScaleX\linewidth\else\Gin@nat@width\fi}

\def\checkGraphicsHeight{\ifdim\Gin@nat@height>.9\textheight
	\tsGraphicsScaleY\textheight\else\Gin@nat@height\fi}

\def\fixFloatSize#1{}
\let\ts@includegraphics\includegraphics

\def\inlinegraphic[#1]#2{{\edef\@tempa{#1}\edef\baseline@shift{\ifx\@tempa\@empty0\else#1\fi}\edef\tempZ{\the\numexpr(\numexpr(\baseline@shift*\f@size/100))}\protect\raisebox{\tempZ pt}{\ts@includegraphics{#2}}}}

\AtBeginDocument{\def\includegraphics{\@ifnextchar[{\ts@includegraphics}{\ts@includegraphics[width=\checkGraphicsWidth,height=\checkGraphicsHeight,keepaspectratio]}}}

\DeclareMathAlphabet{\mathpzc}{OT1}{pzc}{m}{it}

\def\URL#1#2{\@ifundefined{href}{#2}{\href{#1}{#2}}}

\def\UrlOrds{\do\*\do\-\do\~\do\'\do\"\do\-}%
\g@addto@macro{\UrlBreaks}{\UrlOrds}

\edef\fntEncoding{\f@encoding}

\makeatother

\newif\ifmultipleabstract\multipleabstractfalse%
    \makeatletter
\def\ead{\@ifnextchar[{\@uad}{\@ead}}
\gdef\@ead#1{\bgroup
   \def\_{\string\underscorechar\space}
   \def\{{\string\lbracechar\space}
   \def\textdagger{\string\textdagger\space}
   \def\texttildeapprox{\string\texttildeapprox\space}
   \def~{\hashchar\space}
   \def\}{\string\rbracechar\space}
   \edef\tmp{\the\@eadauthor}
   \immediate\write\@auxout{\string\emailauthor
     {#1}{\expandafter\strip@prefix\meaning\tmp}}
  \egroup
}
\gdef\emailauthor#1#2{\stepcounter{ead}
      \g@addto@macro\@elseads{\raggedright
      \let\corref\@gobble
      \eadsep\texttt{#1} (#2)
      \def\eadsep{\unskip,\space}}
}

\makeatother
  
\begin{document}

\begin{frontmatter}

    \title{
  On the selection and effectiveness of pseudo-absences for species distribution modeling with deep learning    
} 
    
\author[a3053a9616d83]{Robin Zbinden\corref{c-40c753e77c82}}
\ead{robin.zbinden@epfl.ch}\cortext[c-40c753e77c82]{Corresponding author.}
\author[a3053a9616d83]{Nina van Tiel}
\author[a6fd8d792e0f6]{Benjamin Kellenberger}
\author[a3053a9616d83]{Lloyd Hughes}
\author[a3053a9616d83]{Devis Tuia}

\address[a3053a9616d83]{
    Ecole Polytechnique Fédérale de Lausanne (EPFL), Environmental Computational Science and Earth Observation (ECEO) Laboratory, Switzerland}
  	
\address[a6fd8d792e0f6]{
    Yale University, Jetz Lab, USA}

\begin{abstract}
Species distribution modeling is a highly versatile tool for understanding the intricate relationship between environmental conditions and species occurrences. However, the available data often lacks information on confirmed species absence and is limited to opportunistically sampled, presence-only observations. To overcome this limitation, a common approach is to employ pseudo-absences, which are specific geographic locations designated as negative samples. While pseudo-absences are well-established for single-species distribution models, their application in the context of multi-species neural networks remains underexplored. Notably, the significant class imbalance between species presences and pseudo-absences is often left unaddressed. Moreover, the existence of different types of pseudo-absences (e.g., random and target-group background points) adds complexity to the selection process. Determining the optimal combination of pseudo-absences types is difficult and depends on the characteristics of the data, particularly considering that certain types of pseudo-absences can be used to mitigate geographic biases. In this paper, we demonstrate that these challenges can be effectively tackled by integrating pseudo-absences in the training of multi-species neural networks through modifications to the loss function. This adjustment involves assigning different weights to the distinct terms of the loss function, thereby addressing both the class imbalance and the choice of pseudo-absence types. Additionally, we propose a strategy to set these loss weights using spatial block cross-validation with presence-only data. We evaluate our approach using a benchmark dataset containing independent presence-absence data from six different regions and report improved results when compared to competing approaches.
\end{abstract}
    
  \end{frontmatter}
  
\section{Introduction}

In a world where climate change and human activities increasingly threaten numerous species and their habitat, there is a growing need to understand the factors that determine the presence of a species at a specific geographic location. Addressing this need,  species distribution models (SDMs) seek to unveil the complex relationship between environmental conditions at a given location and the likelihood of a species occurring there~\citep{elith2009species, franklin2010mapping}. In particular, SDMs are used to predict the geographic range of species, thereby playing a pivotal role in supporting conservation and restoration policies~\citep{guisan2013predicting, sofaer2019development}.
Developing reliable SDMs, however, presents several challenges, such as limited data availability and inherent selection bias within the observations used~\citep{beck2014bias, mesaglio2021overview}. One typical issue is the potential geographic bias arising from variations in sampling efforts across different areas~\citep{kadmon2004effect}. These limitations not only hinder the predictive accuracy of these models, but also impede their ability to generalize effectively to other regions~\citep{elith2010art}. The scarcity of records for rare species further diminishes the significance of model predictions. 

Nevertheless, recent initiatives in community science, exemplified by platforms like \textit{iNaturalist}\footnote{\href{https://www.inaturalist.org/}{https://www.inaturalist.org/}}, \textit{eBird}\footnote{\href{https://ebird.org/home}{https://ebird.org/home}}, and \textit{Pl@ntNet}\footnote{\href{https://plantnet.org}{https://plantnet.org}}, have revolutionized the field and consolidated large numbers of species records contributed by enthusiasts and experts alike. The iNaturalist platform, for instance, comprises an impressive collection of over 150 million species observations spanning more than 400,000 species. These rich data resources present tremendous opportunities to significantly enhance the performance of SDMs and address their aforementioned associated challenges~\citep{botella2023geolifeclef2023, teng2023satbird}.
However, they frequently consist of \textit{presence-only} (PO) observations, providing information exclusively about the presence of species, while lacking any corresponding data regarding their absence~\citep{pearce2006modelling, elith2020presence}. This disparity emerges from the difficulty of gathering data on species absence, in contrast to the more opportunistic nature of recording species presence~\citep{franklin2010mapping}. From a machine learning perspective, this constraint implies that only positive samples are at our disposal~\citep{bekker2020learning}.

\begin{figure}
    \centering
    \includegraphics[width=1\linewidth]{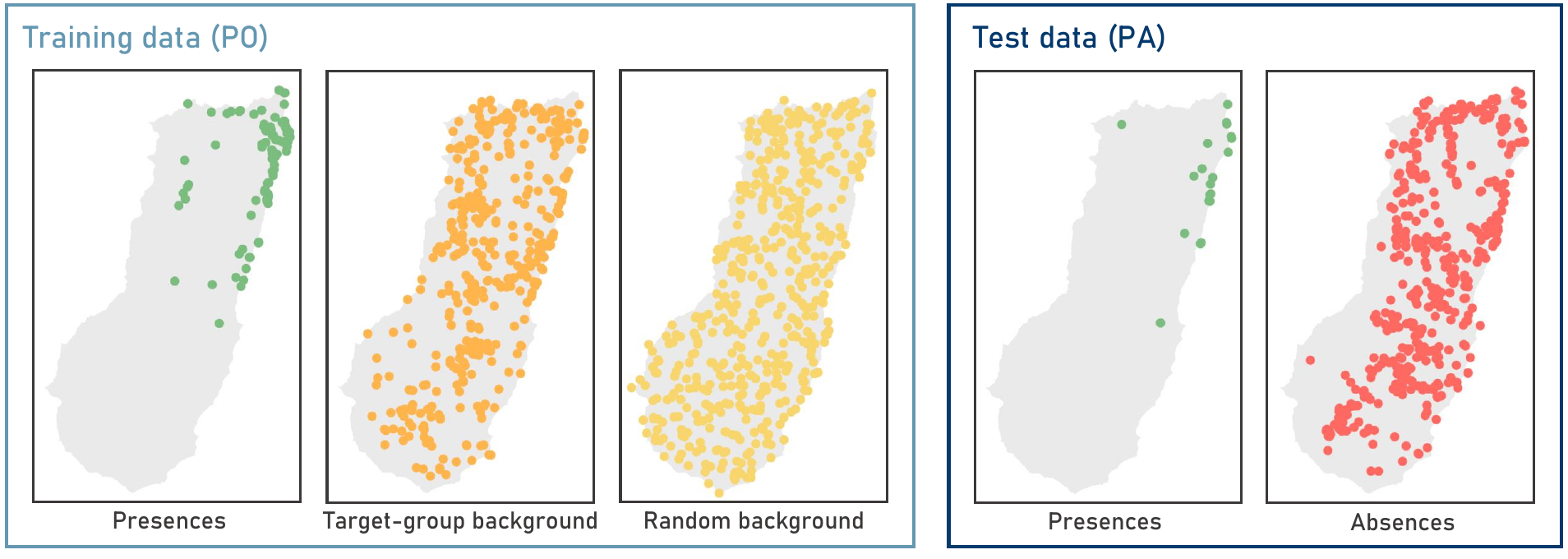}
    \caption{Modeling species distributions often involves utilizing presence-only (PO) data, where information about species absences is unavailable. To apply machine learning techniques in such situations, pseudo-absences are used as a contrast to presence data. There are primarily two types of pseudo-absences: target-group background points and random background points. Target-group background points consist of presences of other species, sharing a similar sampling bias, while random background points are uniformly sampled within the area.
    In this paper, we emphasize the critical importance of how these different types are managed during training for optimal performance, especially when dealing with neural networks. We support our approach by evaluating it on an independent test set comprised of presence-absence (PA) data.
    }
    \label{fig:introduction}
\end{figure}

However, most statistical and machine learning techniques require the incorporation of negative samples to effectively discriminate between species' presence and absence. Different methods have been devised to tackle this challenge, with one common approach being the use of \textit{pseudo-absences} sampling. This technique involves designating selected geographic locations as negative samples, even though there is no guarantee that the environmental conditions at these locations are unfavorable habitats for the target species. These pseudo-absences, also known as \textit{background points} or \textit{pseudo-negatives}, are usually sampled uniformly across the geographic space (\textit{random background points}) or among the presence locations of other species (\textit{target-group background points}), aiming to account for the sampling bias within the presence data~\citep{ponder2001evaluation, phillips2009sample, botella2020bias}. An illustration of these two types of pseudo-absences is provided in Figure~\ref{fig:introduction}.

Incorporating pseudo-absences into model training generally involves combining them with the set of presence data~\citep{valavi2022predictive}. However, this approach raises important questions regarding the optimal \textit{quantity} and \textit{type} of these pseudo-absences, potentially leading to class imbalance issues between presences and pseudo-absences and biased predictions if not carefully managed. While several studies have addressed these questions~\citep{stokland2011species, barbet2012selecting, phillips2009sample}, the answers tend to be context-dependent, making it difficult to establish general guidelines. This challenge is particularly pronounced for models that have not been the focus of these studies, as is the case with \textit{neural networks}, the most popular family of methods of the recent wave of machine learning in ecology~\citep{tuia2022perspectives}.

We aim to bridge this gap since the use of neural networks and \textit{deep learning} techniques stands as a promising and efficient approach for SDMs~\citep{teng2023satbird, botella2023geolifeclef2023, davis2023deep}. These methods have shown significant advancements when applied to large datasets, often outperforming traditional machine learning techniques in various fields~\citep{krizhevsky2012imagenet, christin2019applications, brown2020gpt3, Jumper2021alphafold2}. A distinctive advantage of deep learning approaches lies in their inherent flexibility, enabling the simultaneous integration of diverse data types and the modeling of multiple species, a capability that is increasingly vital in ecological research. Recent research emphasizes the potential of deep learning methods in SDMs, showing performance that matches or exceeds traditional techniques. In particular, this improvement seems to extend to both presence-absence~\citep{chen2017, teng2023satbird} and presence-only datasets~\citep{deneu2021convolutional, botella2023geolifeclef2023}.

In this study, we investigate how to adequately employ pseudo-absences with neural networks. Our emphasis lies on multi-species models, observing that the incorporation of pseudo-absences seamlessly aligns with the multi-label prediction capabilities of neural networks. This integration is achieved through a modification of the training loss function that is used to optimize the neural network parameters.
Specifically, we enhance the \textit{full} loss function proposed by~\cite{cole2023spatial} by assigning appropriate weights within the loss function for presences and pseudo-absences. This adjustment addresses the class imbalance between pseudo-absences and presences and recognizes that some pseudo-absences can be more informative than others in some situations~\citep{phillips2009sample}.
Since our loss function involves weights whose optimal values may vary depending on the dataset, we show how to tune them using presences-only data with spatial block cross-validation~\citep{roberts2017cross}.

We assess the effectiveness of our approach using a well-established benchmark dataset in SDMs, which comprises data from six distinct regions~\citep{elith2020presence}. This dataset includes independent presence-absence data for model evaluation, marking our work as the first to assess different SDMs loss functions on presence-absence data. In addition, previous evaluations have predominantly focused on global scales. While crucial for establishing global biodiversity trends, finer distribution maps are essential for informing local conservation policies. The diverse regions considered in our study span various scales, from local to continental. Our approach not only exhibits superior performance compared to alternative methods for neural networks, especially when coupled with an effective cross-validation methodology, but also demonstrates adaptability to accommodate the specific characteristics of diverse datasets. This includes addressing issues such as class imbalance and sampling bias.

\section{Background}
\label{sec:background}


\subsection{Pseudo-absences in single-species models}

Pseudo-absences were introduced as a means to train single-species models in situations where data on species absence is unavailable, as is the case with presence-only datasets~\citep{pearce2006modelling}.
These pseudo-absences serve the essential purpose of creating a contrast with presence data, thus preventing the model from converging to a trivial solution where it predicts the presence of the species everywhere.
Extensive research has delved into the integration of pseudo-absences in single-species modeling~\citep{Wisz2009, barbet2012selecting, senay2013novel, jarnevich2017minimizing}. These studies predominantly center on fundamental questions related to the required quantity and the appropriate type of pseudo-absences.

\textbf{Quantity of pseudo-absences.}
Conventional approaches sample a specific number of pseudo-absences, typically fixed~\citep{elith2006novel} or relative to the number of available occurrences per species~\citep{valavi2021modelling}. That said, the optimal number often depends on the model in use. For instance, in the case of \textit{Maxent}~\citep{phillips2006maximum}, incorporating a larger number of pseudo-absences is generally more advantageous, up to a saturation point~\citep{phillips2008modeling}. 
For classification models such as \textit{Random Forests}, it is recommended to also sample a large number of pseudo-absences, but then to employ bootstrapping to ensure a balanced count of presences and pseudo-absences for each tree~\citep{barbet2012selecting, valavi2021modelling}.

\textbf{Type of pseudo-absences.}
Diverse methods exist for sampling pseudo-absences. The standard approach involves randomly selecting points within the study area, which we refer to as ``random background points'' throughout this paper. Some alternative approaches restrict the sampling of pseudo-absences to areas geographically distant from known presences~\citep{mateo2010profile, barbet2012selecting}. Another method focuses on the environmental space, where a first habitat suitability map is constructed using presence data only and envelope modeling~\citep{araujo2012uses}, and pseudo-absences are drawn from areas predicted to have low suitability~\citep{engler2004improved}. Nonetheless, several studies have indicated that these methods do not consistently outperform the random background point approach and may even amplify biases originating from the presence data~\citep{Wisz2009, lyu2022integrated}.

Alternatively, the presences of other similar species can be used as pseudo-absences; these are referred to ``target-group background points'' in this context. This approach often shows significant improvement in performance compared to random background points sampling, particularly for datasets marked by substantial sampling bias~\citep{ponder2001evaluation, phillips2009sample}. Several factors contribute to the often superior performance of these pseudo-absences, with a key factor being that target-group background points often incorporate a bias similar to that of the presences of the target species~\citep{hertzog2014field, botella2020bias}. This alignment of biases enables a finer discrimination between presence and absence. 

A visual example of the difference in distribution between the target-group and random background points is represented in the left side of Figure~\ref{fig:introduction}.
In practice, studies rely on either random background points or target-group background points, but seldom combine both. However, combining them may potentially enhance model performance, an aspect we investigate in this work.

\subsection{Pseudo-absences in multi-species neural network models}
\label{sec:background_multi}


In this section, our focus shifts towards modeling the distributions of multiple species simultaneously, often referred to as multi-species distribution models~\citep{poggiato2021interpretations}. In machine learning terminology, this corresponds to a multi-label classification task. This differs from the more conventional multi-class classification, where only one species would be assumed to be present at a given location.
In the context of presence-only data, not only are we restricted to presence records, but we often have only one species occurrence record per location, as exemplified by datasets like iNaturalist. We essentially find ourselves in a situation described as \textit{single positive multi-label learning} by~\cite{cole2021multilabel}. In this setup, we have knowledge of one positive label (observed presence) and unknown labels for (usually all) other species. However, similar to single-species modeling, we still need to employ pseudo-absences to prevent the model from incorrectly predicting the presence of all species at every location. 
The simplest way to approach pseudo-absences in this case is to assume that the missing labels are absences. This approach, known as ``assume negative'', corresponds to the definition of target-group background points. As target-group background points alone may not adequately cover the entire geographic area, it can be beneficial to include additional random background points. In the context of neural networks, an effective way to combine these two types of pseudo-absences is by adjusting the training loss function.

In the field of machine learning, training a model often involves minimizing a loss function that quantifies the model's error on training data. Generally, deep learning-based SDMs are trained with the binary cross-entropy loss~\citep{benkendorf2020effects, deneu2021convolutional, zhang2022novel, zbindenspecies}, with the target-group background points often used as the \textit{de facto} negative samples when absence data is unavailable. Nevertheless, the loss function can be modified to reflect the specificities of the problem. In our case, we can adapt the loss function to inform the model about how pseudo-absences are integrated. This can be achieved by adjusting or assigning weights to the different components of the loss function.
Recently,~\cite{cole2023spatial} introduced the following \textit{full assume negative loss}, designed to account for both target-group background points and random background points:
\begin{align}
    L_{\text{full}}(\mathbf{y}, \mathbf{\hat{y}}) = - \frac{1}{S} \sum^S_{s=1}\biggl[ \underbrace{\mathbbm{1}_{[y_s=1]} \lambda \log(\hat{y}_s)}_{\text{presences}} + \underbrace{\mathbbm{1}_{[y_s=0]} \log(1 - \hat{y}_s)}_{\text{target-group background}} + \underbrace{\log(1 - \hat{y}_s')}_{\text{random background}} \biggr]
\end{align}
where $y_s$ is $1$ if species $s$ is present (the species has been observed) and $0$ otherwise, $\hat{y}_s$ denotes the predicted suitability score for species $s$ (ranging from 0 to 1), $\hat{y}_s'$ represents the model's prediction for species $s$ at a random location, $S$ is the number of species considered, and $\mathbbm{1}_{[\cdot]}$ is the indicator function, returning $1$ if the condition inside the brackets is true and $0$ otherwise.
It is important to note that this loss function is defined for a single presence location, and therefore the overall loss for the entire training dataset is computed by averaging over all the presence locations in the training set.
We observe that $L_{\text{full}}$ is composed of three distinct terms, each corresponding to a different type of sample. The first term addresses presences, favoring a high score for the species that is present, and is weighed by a coefficient $\lambda$ to compensate for the relatively small impact of presences in the loss function. The second term represents target-group background points and assumes that species not observed are absent. Finally, the third term corresponds to random background points, where it is assumed that all species are absent. 
Consequently, $L_{\text{full}}$ effectively combines target-group and random background points in a straightforward and clear manner.

\begin{figure}
    \begin{subfigure}[t]{0.361\textwidth}
    \includegraphics[width=1\linewidth]{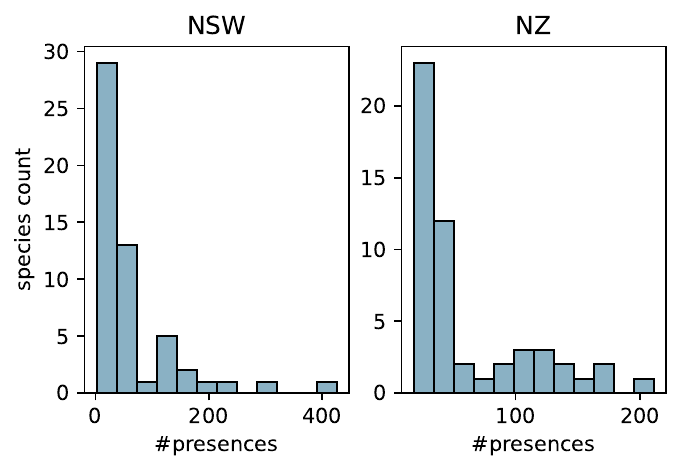} 
    \vspace{-20pt}
    \caption{}
    \label{fig:longtail}
    \end{subfigure}
    \hspace{7.4pt} \unskip\ \vrule\ \hspace{7.4pt}
    \begin{subfigure}[t]{0.6\textwidth}
    \includegraphics[width=1\linewidth]{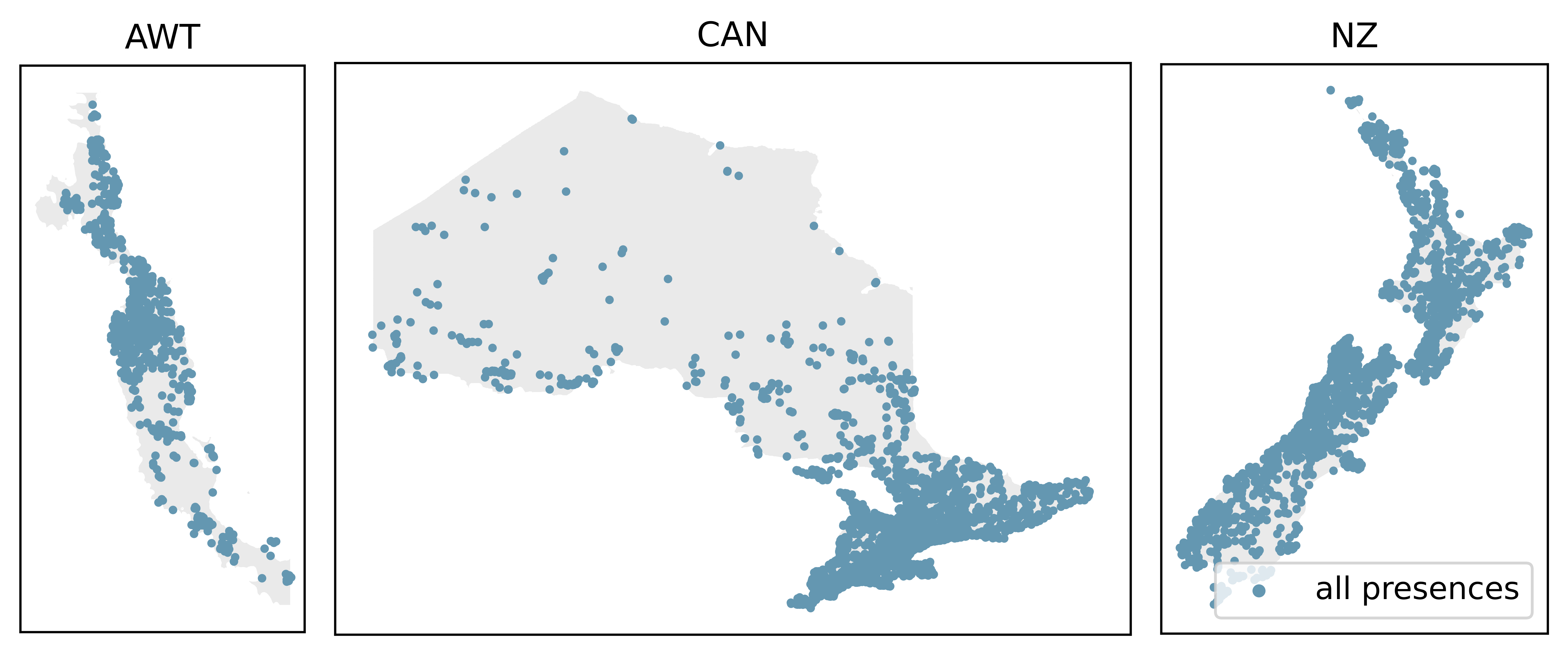}
    \vspace{-20pt}
    \caption{}
    \label{fig:bias}
    \end{subfigure}
    
    \caption{Species occurrence records generally exhibit sampling biases, as depicted here by training presences in the dataset from~\cite{elith2020presence}. (a) The number of presences per species follows a long-tailed distribution, with many species having only a limited number of available observations. To address this issue, we incorporate a species weight $w_s$ for each species in our loss function. (b) The geographic distribution of the presence records of all species shows varying biases across regions. We introduce the pseudo-absence weight $\lambda_2$ to mitigate this bias. 
    Additional plots for the remaining regions, not presented here, can be found in \ref{sec:dataappendix}.}
    \label{fig:distribution}
\end{figure}

\textbf{Limitations of $L_{\text{full}}$.} Primarily, this loss function fails to adequately address the issue of class imbalance within datasets. Two key aspects of class imbalance need to be tackled here: 
\begin{enumerate}
    \item There is often a disparity between the quantity of pseudo-absences required and the number of presences per species in typical datasets, often with a larger number of pseudo-absences. This imbalance can impede the effective training of machine learning models~\citep{johnson2019classimbalance} and has been shown to lead to suboptimal results for SDMs~\citep{benkendorf2023correcting}. While the introduction of the weight $\lambda$, whose value is usually proportional to the number of species~\citep{mac2019presence, cole2023spatial}, aims to alleviate this issue, it does not address the second aspect described below.
    \item We have to ensure that each species is equally weighted within the loss function. The uniformity of $\lambda$ across all species fails to account for variations in the number of presences among species. For example, Figure~\ref{fig:longtail} shows that a significantly higher number of reported presences is available for some species, which results in a long-tailed distribution. In their work,~\cite{cole2023spatial} address this issue by downsampling the presences for species with a large number of occurrences. However, this approach underutilizes available data, potentially leading to a decline in performance. 
\end{enumerate}

In addition to the class imbalance issue, the target-group background points and random background points are assigned the same weight within $L_{\text{full}}$. However, empirical studies suggest that target-group background points typically provide more valuable information than random ones~\citep{phillips2009sample, botella2020bias} when there is a significant bias in the presence data, as it is often the case with presence-only data (see Figure~\ref{fig:bias} and~\cite{beck2014bias, mesaglio2021overview}). Ideally, the loss function should offer flexibility in determining the proportion of target-group background points relative to random background points, since this proportion is dependent on the specific dataset under consideration. Our method tackles all these aspects and proposes a new loss function, at the same time accounting for species observations imbalance, different types of pseudo-absences and their relative proportions.

\section{Methods}
\label{sec:methods}

We introduce a new loss function for SDMs, described in Section~\ref{sec:fullweightedloss}. This loss function incorporates dataset-dependent weights that get tailored to the specific characteristics of the data. In Section~\ref{sec:cv}, we advocate for the use of block cross-validation to determine the optimal values of these weights. While the loss function is applicable to any deep learning method, we employ in our experiments a particular multi-species neural network model that is detailed in Section~\ref{sec:model}. Subsequently, we provide an overview of the dataset in use in Section~\ref{sec:dataset}, and elaborate on our evaluation methodology in Section~\ref{sec:baselines}.

\subsection{Full weighted loss function}
\label{sec:fullweightedloss}

In light of the considerations discussed above, we propose an extension of $L_{\text{full}}$. We refer to this new loss function as the \textit{full weighted loss function}, defined as follows:
\begin{align}
    L_{\text{full-weighted}}(\mathbf{y}, \mathbf{\hat{y}}) = - \frac{1}{S} \sum^S_{s=1} \biggl[ \underbrace{\mathbbm{1}_{[y_s=1]} \lambda_1 w_s  \log(\hat{y}_s)}_{\text{weighted presences}}   +  \overbrace{\mathbbm{1}_{[y_s=0]} \lambda_2 \frac{1}{\left(1 - \frac{1}{w_s}\right)}  \log(1 - \hat{y}_s)}^{\text{weighted target-group background}} \nonumber  \\ + \underbrace{(1 - \lambda_2) \log(1 - \hat{y}_s')}_{\text{weighted random background}}\biggr].
\end{align}
The key distinction between $L_{\text{full}}$ and our new formulation lies in the introduction of weighting terms, with the addition of the coefficients $w_s$, $\lambda_1$, and $\lambda_2$. We will now provide a detailed explanation of each of these weights.

\textbf{Species weights.}
To address the class imbalance issues, we introduce the species weights denoted as $w_s$. For each species $s$, the weight is defined as:
\begin{align}
        w_s = \frac{n}{n_{\text{p($s$)}}} = \frac{1}{\text{freq(\textit{s})}}
\end{align}
where $n_{\text{p($s$)}}$ represents the number of presence records for species $s$ and $n$ is the total number of presence locations in the training set. We explore alternative definitions of $w_s$ in Section~\ref{sec:results}. In this formulation, $w_s$ corresponds to the inverse of the frequency of species $s$. Weighting the species' presences this way ensures that the effective contribution of the presence records to the loss is equivalent for every species. Nevertheless, solely applying this weight to presences is insufficient to guarantee the unbiased treatment of all species by the model, as the number of target-group background points may be disproportionally high for species with few observations. Hence, to ensure that their contribution is consistent for every species, we also weight target-group background points with the following term:
\begin{align}
    \frac{1}{\left(1 - \frac{1}{w_s}\right)} = \frac{n}{n_{\text{tgbg($s$)}}} = \frac{1}{1 - \text{freq(\textit{s})}}
\end{align}
where $n_{\text{tgbg($s$)}} = n - n_{\text{p($s$)}}$ denotes the number of target-group background points.
Importantly, by weighting both presences and target-group background points, we guarantee that the individual contributions to the loss function of presences, target-group background points, and random background points in the training set are equivalent. In this way, we avoid assuming that the number of presences in the training set reflects the prevalence of the species, as prevalence cannot be determined solely from presence data~\citep{hastie2013inference}. However, if additional information about prevalence is available, it can be integrated by scaling the species prediction scores \textit{a posteriori}, with the flexibility to scale differently for each species.

\textbf{Pseudo-absence weight.}
We introduce the pseudo-absence weight $\lambda_2$ as a way to indicate which type of pseudo-absence should have more importance in the learning process. Some studies have demonstrated that using target-group background points may lead to more accurate predictions~\citep{phillips2009sample, botella2020bias}, but including a portion of random background points can be useful for covering the entire geographic area of interest~\citep{vanderwal2009selecting, cole2023spatial}. With $\lambda_2 \in [0,1]$, our proposed loss modulates the emphasis between these two types of pseudo-absence. When $\lambda_2 = 0$, only random background points are used, while $\lambda_2 = 1$ means that only target-group background points are employed. The full loss of~\cite{cole2023spatial} is equivalent to $\lambda_2 = 0.5$, assuming an equal weight on target-group background points and random background points.

\textbf{Presence weight.} 
In addition, we introduce the presence weight $\lambda_1$ as a means to adjust the weighting of the presences compared to the pseudo-absences. When $\lambda_1 > 1$, more emphasis is placed on correctly classifying presences, whereas when $\lambda_1 < 1$, the focus shifts to pseudo-absences. Notably, considering that presence data should be more reliable than pseudo-absences, a relatively higher value of $\lambda_1$ can prove advantageous as it instructs the model to prioritize correctly classifying presences. This weight also makes $L_{\text{full-weighted}}$ a generalization of $L_{\text{full}}$, enabling direct comparison. $L_{\text{full}}$ is recovered by setting $\lambda_1 = S/2$, $\lambda_2 = 0.5$, and removing the species weights $w_s$.

\subsection{Tuning loss weights}
\label{sec:cv}

Our full weighted loss function includes the weights $\lambda_1$ and $\lambda_2$, whose optimal values are data-dependent and, as hyperparameters, require careful tuning. In machine learning, hyperparameter values are typically determined through cross-validation, involving the partitioning of the training data and designating a portion as the validation set. The selection of hyperparameter values, or equivalently the selection of the model, is then based on the performance on this validation set. However, constructing an effective validation set with presence-only data is challenging, as evaluating model performance with such data may lead to choosing a model that makes biased predictions~\citep{el2018improved}.
To alleviate this issue, we employ spatial block cross-validation~\citep{roberts2017cross, valavi2023flexible}, which involves spatially splitting the presence observations into training and validation sets. This approach makes it difficult for the model to perform well on the validation set, favoring models capable of generalizing to unseen areas~\citep{smith2021validation}. This is particularly important since presence-only data often exhibits biases toward specific areas, and we need to assess the model's ability to counter them.

\begin{figure}
    \centering
    \includegraphics[width=1\linewidth]{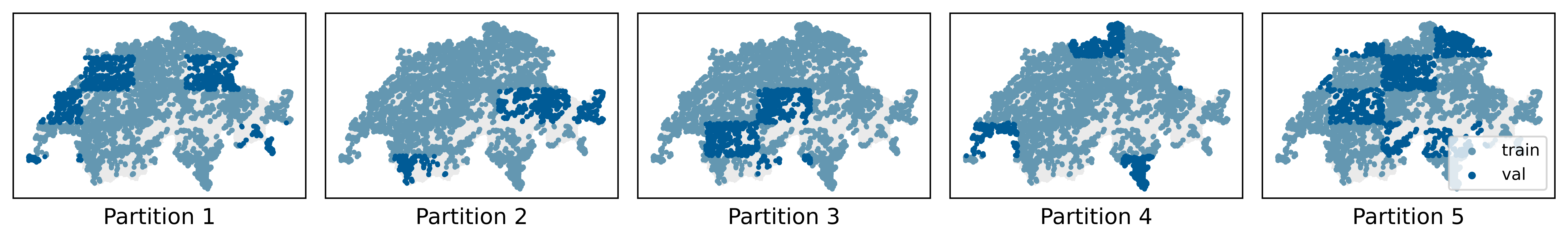}
    \caption{$k$-fold block cross-validation~\citep{roberts2017cross} is used to find the optimal value of the pseudo-absence weight $\lambda_2$. It involves the spatial partitioning of presence observations into the training and validation sets, comprising respectively 80\% and 20\% of the samples. The presence records considered here pertain to the Swiss region of the dataset described in Section~\ref{sec:dataset}.}
    \label{fig:blockcv}
\end{figure}

Hence, we perform $k$-fold block cross-validation, dividing the region of interest into $5 \times 5$ geospatial blocks and assigning them to $k$ distinct folds, such that each fold contains approximately the same number of presence observations. We choose $k=5$, resulting in that for each partition, the model is trained on 80\% of the presences and validated on the remaining 20\%. This procedure is equivalent to the spatial blocks based on rows and columns of~\cite{valavi2018blockcv}, and is illustrated in Figure~\ref{fig:blockcv}. The model is then evaluated on each fold and the results are averaged. The model with the best performance, determined here by the mean AUC (Area Under the receiver operating characteristic Curve) over all the species, is selected. The target-group background points are used as pseudo-absences to compute the AUC on the validation set, similarly to~\cite{valavi2023flexible}.

\subsection{Model and training details}
\label{sec:model}

We adopt a configuration similar to the multi-species model presented in~\cite{zbindenspecies}. Instead of developing individual models for each species, we train a single \textit{multi-layer perceptron} (MLP) model for each region (details about the six regions considered are given in Section~\ref{sec:dataset}). These models predict suitability scores for every species within the region following a multi-label classification approach. MLPs are a type of neural network consisting of multiple layers of fully-connected neurons with non-linear activation functions between layers~\citep{gorishniy2021revisiting}, enabling them to capture complex interactions among the input environmental covariates. This architecture is well-suited for the tabular format of the dataset in use (see Section~\ref{sec:dataset}).
For the model and training hyperparameters, we stick to standard values~\citep{mac2019presence, gorishniy2021revisiting}, while also incorporating recent advancements in deep learning to enhance performance. Specifically, our MLP architecture consists of four layers, each containing 512 neurons and connected with residual connections~\citep{he2016resnet}. We employ batch normalization~\citep{ioffe2015batch} and the Rectified Linear Unit (ReLU) activation function in all layers except the final one, where instead a sigmoid function is used to enable multi-label classification.
The model is trained with a batch size of $256$ for $30$ epochs using the AdamW optimizer~\citep{loshchilov2017decoupled}. Both the weight decay and learning rate are set to $0.0001$. Additionally, we employ a learning rate scheduler with exponential decay of $0.95$, and introduce dropout~\citep{srivastava2014dropout} with a probability of $0.01$.

\subsection{Dataset}
\label{sec:dataset}

We use the benchmark dataset from~\cite{elith2020presence}, which comprises occurrence data for 226 anonymized species from six regions around the world: Australian Wet Tropic (AWT), Canada (CAN), New South Wales (NSW), New Zealand (NZ), South America (SA), and Switzerland (SWI). Recently made public, this dataset has already been employed to evaluate and compare different methods~\citep{elith2006novel, valavi2022predictive}, as well as for conducting analyses of target-group background points~\citep{phillips2009sample}. The training set consists of opportunistically sampled presence-only data, while the test set is built with full presence-absence data for each species (Figure~\ref{fig:introduction}, right).
As predictors, the dataset contains a varying set of 10 to 13 environmental covariates for each region, including both climatic and pedological variables (see~\cite{elith2020presence} for a full description).

This dataset is also representative of the biases and challenges often encountered when training SDMs. Notably, the number of presences per species follows a long-tail distribution, as illustrated in Figure~\ref{fig:longtail}. Additionally, the geographic distribution is often biased, particularly in certain areas of some regions. For instance, as observed in Figure~\ref{fig:bias} and quantified by~\cite{phillips2009sample}, there is a substantial bias in the distribution of presences for the AWT and CAN regions, while NZ exhibits less bias. Finally, some regions encompass species belonging to different biological groups, such as plants or birds. While previous studies used only presences of species within the same group as target-group background points~\citep{ponder2001evaluation, phillips2009sample}, for simplicity we utilize all species within each region to train our models. As a result, we have a total of six models, one for each region.

\subsection{Evaluation and baselines}
\label{sec:baselines}

We compare our approach with various other prediction methods and loss functions. Firstly, we examine the different loss functions proposed by~\cite{cole2023spatial}. The $L_{\text{full}}$ presented in Section~\ref{sec:background_multi} is a combination of two losses: the SSDL (same species, different location) loss, which exclusively uses the random background points and presences, and the SLDS (same location, different species) loss, which only uses the target-group background points and presences. Notably, our full weighted loss generalizes these three loss functions by appropriately setting the $\lambda_1$ and $\lambda_2$ weights. Specifically, $L_{\text{full-weighted}}$ is equal to the SSDL loss when $\lambda_1 = 1$ and $\lambda_2 = 0$, the SLDS loss corresponds to $\lambda_1 = 1$ and $\lambda_2 = 1$, and the full loss is equivalent when $\lambda_1 = S/2$ and $\lambda_2 = 0.5$. In each case, no species weights are applied. We then train the multi-species model defined in Section~\ref{sec:model} on these loss functions. 

Additionally, we include results for the Maxent and Boosting Regression Trees models from~\cite{phillips2009sample}, which were trained on the same dataset, but with target-group background points only. Maxent is the conventional approach to SDMs~\citep{phillips2006maximum}, while the Boosting Regression Trees model represents one of the existing tree-based approaches that perform well in modeling species distributions~\citep{valavi2021modelling}. However, both these approaches necessitate the creation of one model per species, resulting in the management of 226 independent models.

All the different methods are evaluated on the presence-absence test set by computing the Area Under the receiver operating characteristic Curve (AUC) for each species and then calculating the mean across all species per region. We opt for AUC due to its widespread use in SDMs and its general high agreement with independent testing data~\citep{konowalik2021evaluation}, as well as to avoid binarizing predictions. For completeness, we still compute the correlation and the area under the precision-recall gain curve~\citep{flach2015precision}, as done in~\cite{valavi2022predictive}, and include these metrics in the Appendix (see Table~\ref{tab:comparison_cor} and Table~\ref{tab:comparison_aucprg}). Finally, we report the average performance over ten different random seeds to add statistical significance.

\section{Results and Discussion}
\label{sec:results}

We first compare our approach to the other baselines, and then perform ablation studies to understand the role played by every weight in the loss function.

\subsection{Comparison with other approaches}

\begin{table}[]
\centering
\caption{Comparison of the mean AUC over the species of our approach to other works. Our loss generalizes the losses proposed by \citep{cole2023spatial}.
The best mean AUC for each column, achieved across the different loss functions, is highlighted in bold, whereas the best mean AUC obtained from single-species models is underlined.}

\setlength{\tabcolsep}{0.46\tabcolsep}  
\renewcommand{\arraystretch}{1.2} 
\begin{tabular}{lccccccc}
                                                                          & AWT   & CAN   & NSW   & NZ    & SA    & SWI   & \textbf{avg}   \\ 
    
    \hline \hline
    \textbf{\cite{cole2023spatial} losses} \\
    SSDL loss, i.e., $\lambda_1 = 1$ and $\lambda_2=0$                    & 0.619 & 0.533 & 0.619 & 0.698 & 0.709 & 0.796 & 0.662 \\
    SLDS loss, i.e., $\lambda_1 = 1$ and $\lambda_2=1$                    & 0.620 & 0.713 & 0.644 & 0.69 & 0.722 & \textbf{0.839} & 0.705 \\
    Full loss, i.e., $\lambda_1 = S/2$ and $\lambda_2=0.5$                & 0.698 & 0.673 & \textbf{0.723} & \textbf{0.742} & 0.814 & 0.836 & 0.748 \\
    \hline
    \textbf{Full weighted loss (ours)} \\
    $\lambda_1=1 $ and $ \lambda_2=0.8$, with $w_s$                                   & \textbf{0.704} & 0.696 & 0.719 & 0.741 & \textbf{0.815} & 0.836 & 0.752 \\
    $\lambda_1=1 $ and fine-tuned $\lambda_2$, with $w_s$                             & \textbf{0.704} & \textbf{0.714} & 0.719 & 0.741 & \textbf{0.815} & 0.838 & \textbf{0.755} \\
    \hline \hline
    \textbf{Single-species \citep{phillips2009sample}} \\
    Maxent                                                                & \underline{0.732} & 0.716 & \underline{0.741} & 0.738 & \underline{0.798} & 0.837 & \underline{0.760} \\
    BRT                                                                   & 0.700 & \underline{0.728} & 0.738 & \underline{0.740} & 0.792 & \underline{0.842} & 0.757 \\
    \hline
\end{tabular}
\vspace{0.1cm} \\

\label{tab:comparison}
\end{table}

The full weighted loss $L_{\text{full-weighted}}$ is compared to the methods and loss functions in Table~\ref{tab:comparison} on the test set of the ~\cite{elith2020presence} dataset. We evaluate the loss functions introduced in~\cite{cole2023spatial} and observe that the use of target-group background points (SLDS loss) consistently yields superior results compared to using random background points (SSDL loss). Moreover, considering both types of pseudo-absences and increasing the weight assigned to presences further enhances performance (Full loss). In comparison, our approach, which incorporates species weights and where the pseudo-absence weight $\lambda_2$ is set to $0.8$ to put more weight on the target-group background points, achieves slightly superior results on average. Moreover, fine-tuning the value $\lambda_2$ through block cross-validation leads to additional improvements in the CAN region. Notably, these results are achieved without the need for specific $\lambda_1$ values. This superior performance is also visible in the correlation (see Table~\ref{tab:comparison_cor}) and the area under the precision-recall gain curve (see Table~\ref{tab:comparison_aucprg}) metrics.
Ultimately, the single-species models exhibit a performance range comparable to that of our approach but still showcase a slight advantage.

\subsection{Loss weights}

We next conduct ablation studies to understand the role of each weight in our loss function. We adjust each of the three weights in question, one at a time, and keep the other weights at their base values, which are $\lambda_1 = 1$, $\lambda_2 = 0.8$, and $w_s = 1/\text{freq}(s)$.

\begin{table}[]
\centering
\caption{Different values for the species weights $w_s$, with $\lambda_1=1$ and $\lambda_2=0.8$. $w_s$ is required to address the strong class imbalance between presences and pseudo-absences. The clamp operation here restricts low frequencies to a minimum threshold of $0.25$. The best mean AUC obtained for each column is highlighted in bold.}

\setlength{\tabcolsep}{0.46\tabcolsep}  
\renewcommand{\arraystretch}{1.2} 
\begin{tabular}{l|cccccc|c}
                                                & AWT   & CAN   & NSW   & NZ    & SA    & SWI   & avg   \\ 
    \hline \hline
    no species weights                          & 0.653 & \textbf{0.719} & 0.652 & 0.702 & 0.741 & \textbf{0.840} & 0.718 \\
    $w_s = 1/\text{freq(}s\text{)}$             & \textbf{0.704} & 0.696 & \textbf{0.719} & \textbf{0.741} & \textbf{0.815} & 0.836 & \textbf{0.752} \\
    $w_s = 1/\text{clamp(freq(}s\text{))}$       & 0.678 & 0.709 & 0.689 & 0.727 & 0.791 & \textbf{0.840} & 0.739 \\
    $w_s = 1/\sqrt{\text{freq(}s\text{)}}$      & 0.697 & 0.704 & 0.710 & 0.737 & 0.799 & 0.839 & 0.748 \\
    \hline
\end{tabular}

\label{tab:speciesweight}
\end{table}

\begin{figure}
    \centering
    \includegraphics[width=1\linewidth]{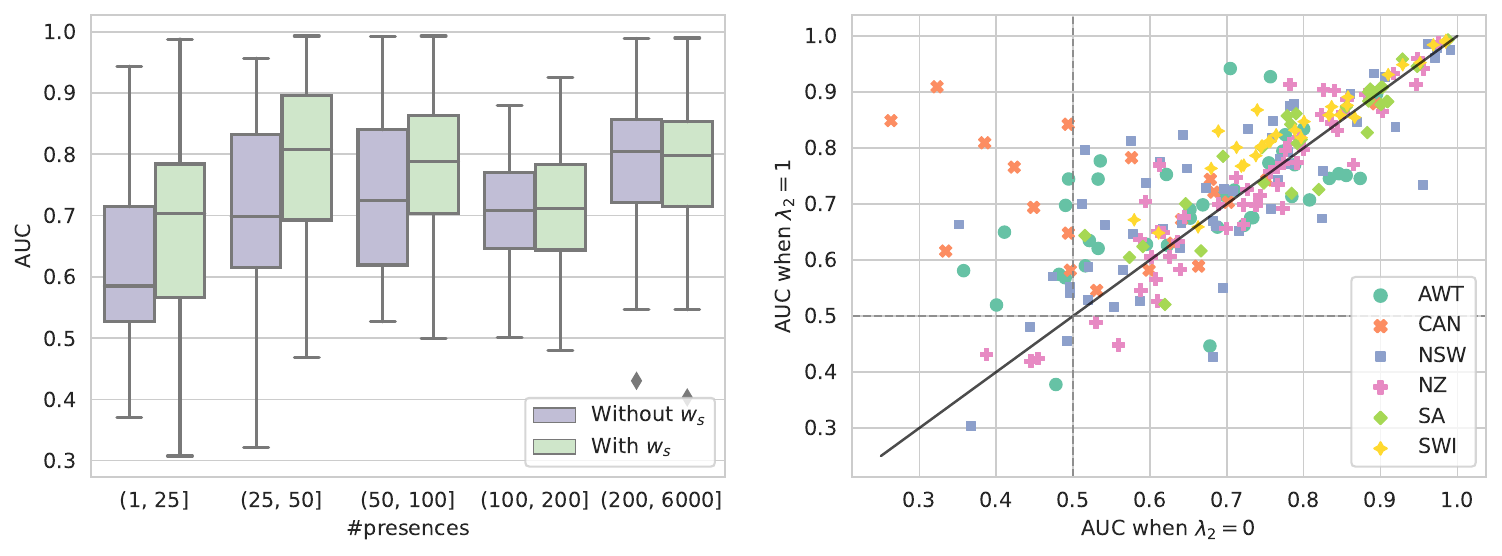}
    \caption{\textbf{Left}: Impact on the AUC when using the species weight $w_s$ in the loss function, grouped by the number of presences records in the training set. The gain of employing $w_s$ is more pronounced for species with fewer presence records. \textbf{Right}: Impact on the AUC when using random ($\lambda_2 = 0$) or target-group ($\lambda_2$ = 1) background points, with every symbol representing a species. While many species benefit from using only target-group background points, not all do.}
    \label{fig:analysis}
\end{figure}

\textbf{Species weights.} We start by assessing the impact of the species weights $w_s$, as presented in Table~\ref{tab:speciesweight}. We compare its influence to a scenario where no species weights are considered and also test different values of $w_s$. Our observations reveal a notable improvement when incorporating species weights, with positive effects evident in four out of six regions, resulting in a higher overall average improvement. Particularly, regions NSW and NZ show substantial increases, effectively addressing the challenge posed by the long-tail distribution of presences for these two regions (see Figure~\ref{fig:longtail}). Alternative approaches involving the square root or clamping low frequency at $0.25$ produce comparable, albeit slightly less accurate results on average.
Furthermore, the left panel of Figure~\ref{fig:analysis} illustrates that incorporating species weights is especially advantageous for species with fewer presence records, with diminishing benefits as the number of presence records increases.

Finally, with the inclusion of the species weights, the addition of the presence weight $\lambda_1$ to the presences, as in $L_{\text{full}}$, is no longer necessary, as demonstrated by the results in Table~\ref{tab:lambda1} in the Appendix. Indeed, the improvements resulting from varying the value of $\lambda_1$ become only marginal.

\textbf{Pseudo-absence weight.} In Table~\ref{tab:lambda2}, we explore the impact of varying the ratio of target-group background points to the pseudo-absences using the weighting parameter $\lambda_2$. As previously observed in Table~\ref{tab:comparison}, the incorporation of target-group background points consistently leads to improvements, extending the observation of~\cite{phillips2009sample} to neural networks. When examining the overall performance across datasets, employing either only target-group background points ($\lambda_2 = 1$) or a percentage of 80\% ($\lambda_2 = 0.8$) tends to produce the best outcomes, although the optimal proportion varies from one region to another. Specifically, for the three regions NSW, NZ, and SA, performance improvements level off for $\lambda_2$ values exceeding $0.6$. This suggests that the ideal ratio of target-group background points is dependent on the specific data characteristics of each region. In particular, this dependency is linked to the sampling bias, as quantified by~\cite{phillips2009sample}. Regions such as AWT, CAN, and SWI, which exhibit more pronounced sampling bias, benefit more from the inclusion of target-group background points compared to other regions. Consequently, employing a block cross-validation procedure to ascertain the optimal value of $\lambda_2$ for each region proves to be beneficial, as visible in the final row of Table~\ref{tab:lambda2}. This procedure identifies the optimal $\lambda_2$ value for three out of the six regions, with the performances of the remaining regions closely approaching the best mean AUC achievable. This leads to a higher average performance across the regions.

\begin{table}[]
\centering
\caption{Varying the value of the pseudo-absence weight $\lambda_2$ has a notable impact on the mean AUC over the species. The results presented in the last row are obtained using block cross-validation, where $\lambda_2$ is selected based on the performance on the validation set.
In each result, the species weight $w_s$ is employed, while $\lambda_1$ is fixed to 1. The best mean AUC in each column is highlighted in bold.}

\setlength{\tabcolsep}{0.46\tabcolsep}  
\renewcommand{\arraystretch}{1.2} 
\begin{tabular}{l|cccccc|c}
                        & AWT   & CAN   & NSW   & NZ    & SA    & SWI   & avg   \\ \hline \hline
$\lambda_2 $=0          & 0.651 & 0.547 & 0.685 & 0.731 & 0.804 & 0.794 & 0.702 \\
$\lambda_2 $=0.2        & 0.673 & 0.600 & 0.710 & 0.740 & 0.812 & 0.814 & 0.725 \\
$\lambda_2 $=0.4        & 0.691 & 0.634 & 0.720 & \textbf{0.743} & 0.815 & 0.824 & 0.738 \\
$\lambda_2 $=0.6        & 0.702 & 0.665 & \textbf{0.722} & \textbf{0.743} & \textbf{0.816} & 0.831 & 0.746 \\
$\lambda_2 $=0.8        & \textbf{0.704} & 0.696 & 0.719 & 0.741 & 0.815 & 0.836 & 0.752 \\
$\lambda_2 $=1          & 0.696 & \textbf{0.714} & 0.713 & 0.738 & 0.811 & \textbf{0.838} & 0.752 \\
\hline
$\lambda_2 $ found by cv & \textbf{0.704} & \textbf{0.714} & 0.719 & 0.741 & 0.815 & \textbf{0.838} & \textbf{0.755} \\
 \hline
 
\end{tabular}

\label{tab:lambda2}
\end{table}

In the right panel of Figure~\ref{fig:analysis}, we present the AUCs obtained when using only target-group background points ($\lambda_2 = 1$) versus using only random background points ($\lambda_2 = 0$) for each species. Points above the diagonal black line indicate species where the use of target-group background points leads to a higher AUC than using random background points, and vice versa. Many species, especially in the CAN region, benefit from using target-group background points, with some achieving a boost as high as $0.5$ points AUC. However, this trend is not present for all species; some obtain lower performance when using only target-group background points instead of random background points. In particular, within the same geographic region, species exhibit varied responses, highlighting that the optimal value of $\lambda_2$ can vary even within a single region, depending on the species.

\begin{figure}
    \centering
    \includegraphics[width=1\linewidth]{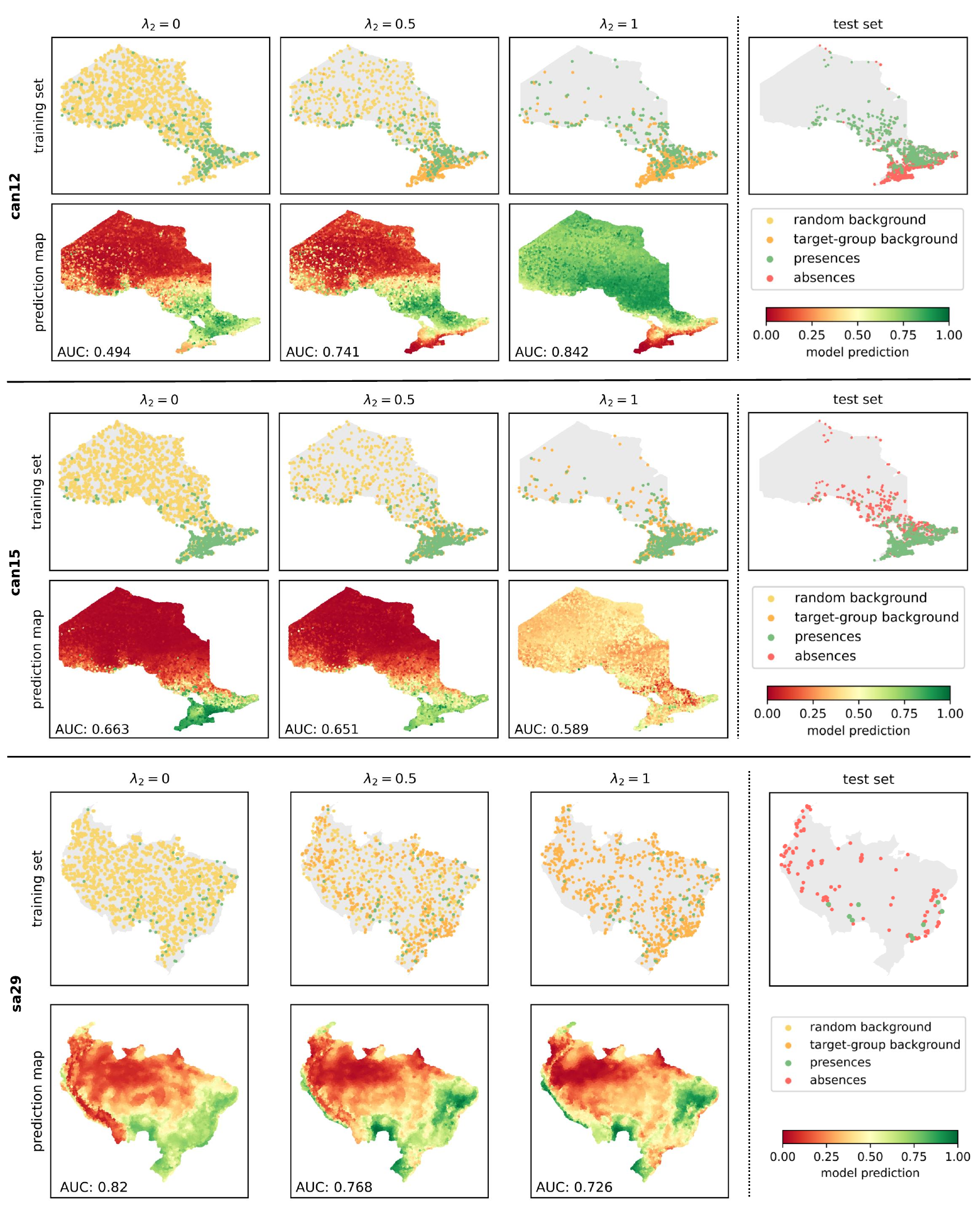}
    \caption{Focus on three distinct (anonymized) species, labeled as \texttt{can12}, \texttt{can15}, and \texttt{sa29}. Each species is represented by its respective training set, test set, and prediction maps. These visualizations illustrate the role and impact of different values for the pseudo-absence weight $\lambda_2$ on the prediction maps generated by the model.}
    \label{fig:predictionmap}
\end{figure}

To gain a concrete understanding of how the choice of $\lambda_2$ influences prediction maps, we present maps for three distinct species, each with varying values of $\lambda_2$—ranging from $0$ (no target-group background points) to $1$ (only target-group background points). The selection of these three species aims at explicitly showcasing the heterogeneous response to $\lambda_2$. Specifically, the more widely distributed species \texttt{can12} from Canada significantly benefits from including target-group background points, whereas the more specialized \texttt{can15} species suffers from relying on target-group background points only. This observation aligns with the findings of~\cite{ranc2017performance} and~\cite{botella2020bias}, which conclude that generalist widely-distributed species gain more from target-group background points compared to specialized species. Notably, the \texttt{can15} species exhibits a distribution closely mirroring the distribution of presences of all species, heavily biased towards the south (as depicted in the panel showing presences and target-group background points only). Consequently, the model struggles to distinguish between target-group background points and presences, resulting in predicted scores hovering around $0.5$ for the entire map. 
Finally, we note that the differences in prediction maps are not always strong, as illustrated by the \texttt{sa29} species. In this case, the distinction lies more in the magnitude of the predictions, although the southeast region is not accurately predicted when using only target-group background points.

\section{Future work}
\label{sec:future_work}

We present several potential extensions to our approach.
Firstly, our loss function offers the flexibility to include as many random background points as desired, as they are appropriately weighted. This flexibility allows one to choose the spatial extent covered by random background points that is optimal~\citep{vanderwal2009selecting}, without being constrained by a specific number. Random background points could also be selectively sampled from subareas within the region of interest, such as those distant from known presences~\citep{mateo2010profile, iturbide2015framework}. Moreover, presence-only data often contains only a single species record per location. In this situation, our approach can be effectively combined with methods that aggregate nearby species presences, such as~\cite{kellenberger2022}.

Furthermore, we observed in Figure~\ref{fig:analysis} (right) and Figure~\ref{fig:predictionmap} that different species within the same region may have different optimal values for the hyperparameter $\lambda_2$. Consequently, setting distinct $\lambda_2$ values per species could improve model performance. In particular, higher values of $\lambda_2$ could be set for known generalist species, especially if there is a strong sampling bias. While the evaluation in this study focused on multi-species neural networks, it is straightforward to adapt our loss function to single-species neural networks by removing the summation over species and retaining only the species of interest. 

We chose to use the benchmark dataset from~\cite{elith2020presence} due to its established reputation in SDMs and its inclusion of PA data to evaluate models trained on PO data. However, the tabular format of this dataset typically reduces the performance of neural networks~\citep{borisov2022deep, grinsztajn2022tree}. Our results align with this limitation, as non-neural network approaches demonstrated comparable or even slightly superior performance compared to our method. Nevertheless, certain studies suggest that deep learning-based SDMs can surpass traditional machine learning methods by incorporating diverse data types~\citep{zhang2022novel, botella2023geolifeclef2023, teng2023satbird}. As a result, we intend to expand this research to datasets involving various types of data, such as environmental rasters, satellite images, or time series data, as exemplified by the GeoLifeCLEF 2023 dataset~\citep{botella2023geolifeclef2023}. An interesting avenue to explore is whether the behavior of the full weighted loss function remains consistent across various neural network architectures, including more complex models such as convolutional neural networks~\citep{krizhevsky2012imagenet} or Transformers~\citep{vaswani2017attention}.

\section{Conclusion}
\label{sec:conclusion}

Employing deep learning methods for SDMs represents a promising approach to processing the substantial volume of new ecological data emerging from community science. The prevalence of presence-only observations, a common format for such datasets, poses a challenge due to the absence of information about the non-occurrence of species. To tackle this fundamental issue, pseudo-absences are frequently employed as contrasting samples to the presence observations.
In this paper, we introduced a unified and flexible approach to integrating different types and quantities of pseudo-absences when using multi-species neural networks. This is achieved through the introduction of a novel loss function tailored to the specific characteristics of the datasets, effectively addressing challenges such as geographic biases and class imbalance issues.
The training of multi-species neural networks with our proposed loss function yielded superior performance compared to previous approaches. Notably, our models performed similarly to Maxent on an independent test set consisting of both presence and absence data from diverse regions while maintaining the advantage of having a single model per region.
Our study also sheds light on the intricate relationship between the type of pseudo-absence and the spatial bias of observations, emphasizing the importance of considering this factor in model development. By providing a comprehensive solution to the incorporation of pseudo-absences, our work opens avenues for further refinement and enhancement of multi-species neural network models, ultimately supporting more accurate and reliable predictions in ecological research and conservation efforts.

\section*{Funding}
This work was supported by the Swiss National Science Foundation, under grant 200021\_204057 ``Learning unbiased habitat suitability at scale with AI (deepHSM)''.

\section*{CRediT authorship contribution statement}
\textbf{Robin Zbinden:} Conceptualization, Methodology, Software, Validation, Formal analysis, Investigation, Data Curation, Writing - Original Draft, Writing - Review \& Editing, Visualization. \textbf{Nina van Tiel:} Conceptualization, Software, Data Curation, Writing - Review \& Editing. \textbf{Benjamin Kellenberger:} Conceptualization, Writing - Review \& Editing, Supervision, Funding acquisition. \textbf{Lloyd Hughes:} Conceptualization, Supervision. \textbf{Devis Tuia:} Conceptualization, Writing - Review \& Editing, Supervision, Project administration, Funding acquisition.

\section*{Declaration of Competing Interest}
The authors declare that they have no known competing financial interests or personal relationships that could have appeared to influence the work reported in this paper.

\section*{Declaration of Generative AI and AI-assisted technologies in the writing process}
During the preparation of this work the authors used GPT-3.5 in order to improve the readability and language of this work. After using this tool, the authors reviewed and edited the content as needed and take full responsibility for the content of the publication.

\newpage
\bibliographystyle{elsarticle-num}
\bibliography{literature}

\newpage
\appendix

\section{Additional Dataset Details}
\label{sec:dataappendix}
\begin{figure}[h]
    \centering
    \includegraphics[width=0.9\linewidth]{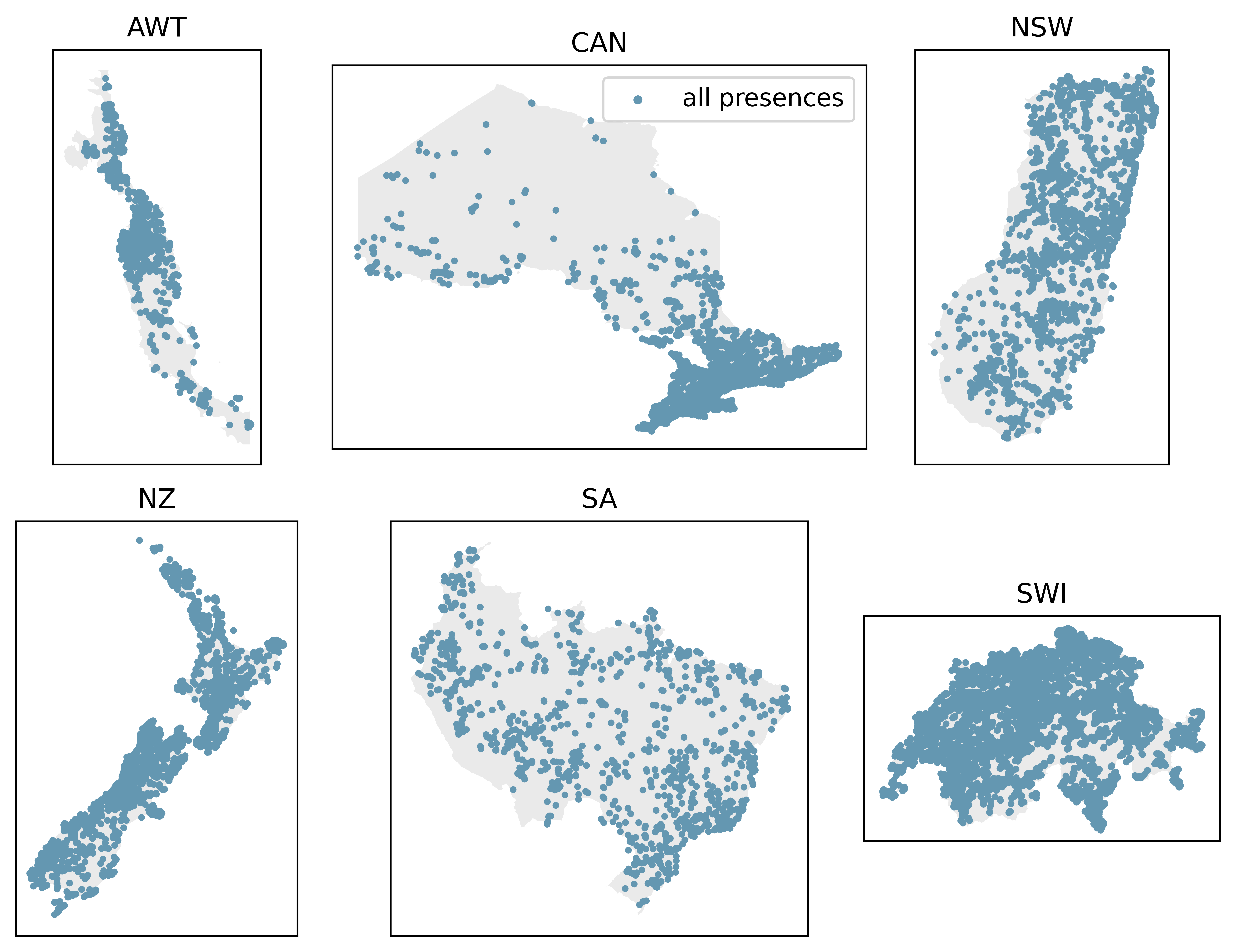}
    \caption{The geographic distribution of the presence records of all species within each region of the dataset in use~\citep{elith2020presence}. The intensity of the sampling bias varies significantly from one region to another.}
    \label{fig:biasall}
\end{figure}

\begin{figure}[h]
    \centering
    \includegraphics[width=1\linewidth]{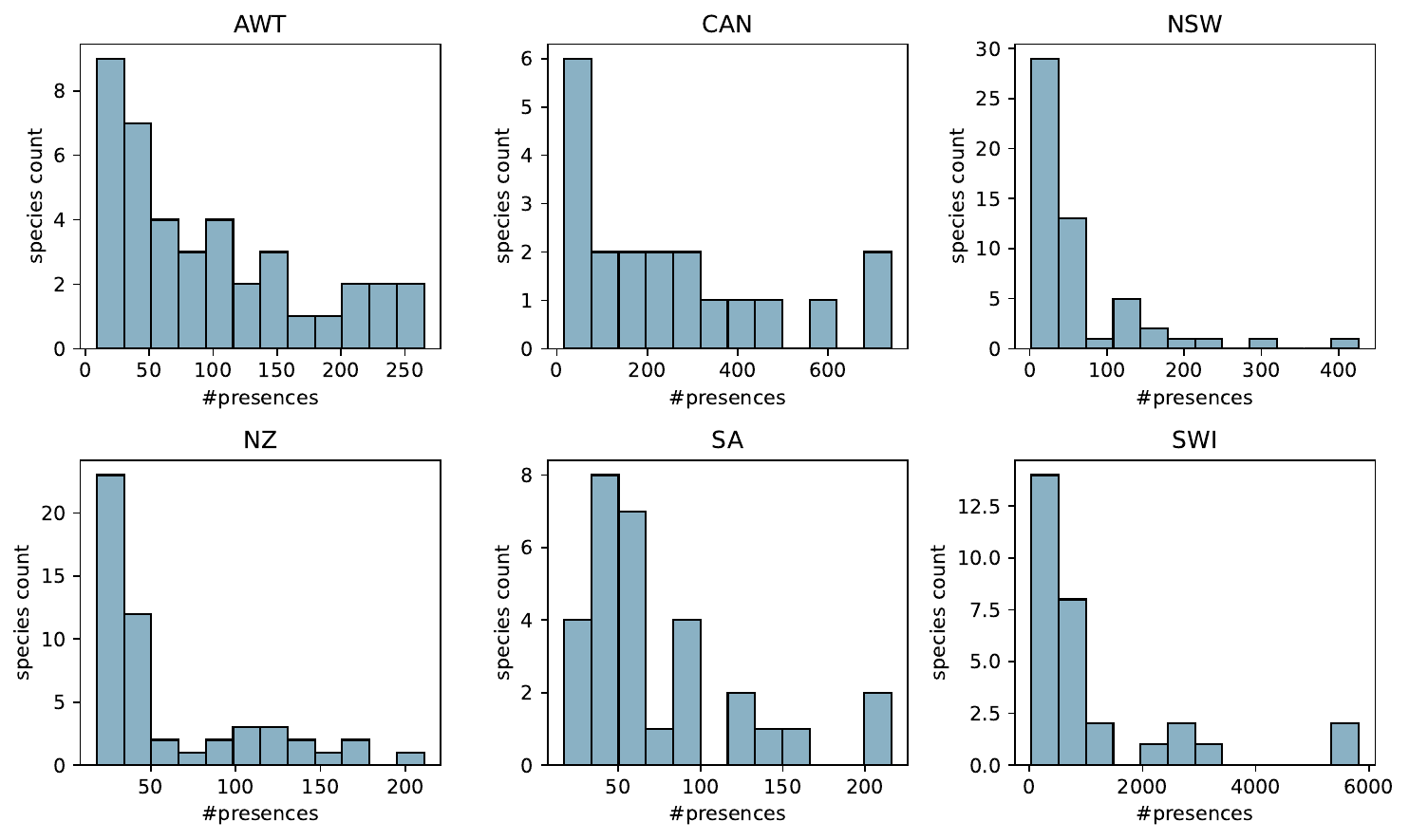}
    \caption{Distribution of the number of presence records per species in each region of the dataset in use~\cite{elith2020presence}. The distributions typically exhibit a long-tailed pattern, with many species observed only a few times, while a few others are observed more frequently.}
    \label{fig:longtailall}
\end{figure}

We provide the maps of the geographic distribution of the presence records for all species (Figure~\ref{fig:biasall}), along with the distribution of the number of presence records per species (Figure~\ref{fig:longtailall}) for all regions. For additional details about the dataset, refer to~\cite{elith2020presence}.

\section{Additional Results}
\label{sec:resultsappendix}

\textbf{Presence weight.} In Table~\ref{tab:lambda1}, we evaluate the impact of the presence weight $\lambda_1$ on performance. Notably, the performance changes from variations in the value of $\lambda_1$ are minor, suggesting that the species weights adequately address the class imbalance between presences and pseudo-absences.

\begin{table}[h]
\centering
\caption{Adjusting the value of the presence weight $\lambda_1$ has a minimal impact on the mean AUC over the species,  indicating that the species weights $w_s$ effectively address class imbalance issues. $\lambda_2$ is fixed at $0.8$ and $w_s$ is used. The best mean AUC in each column is highlighted in bold.}

\setlength{\tabcolsep}{0.46\tabcolsep}  
\renewcommand{\arraystretch}{1.2} 
\begin{tabular}{l|cccccc|c}
                            & AWT   & CAN   & NSW   & NZ    & SA    & SWI   & avg   \\ \hline \hline
$\lambda_1 $=0.1            & \textbf{0.709} & \textbf{0.701} & 0.701 & 0.734 & 0.784 & 0.837 & 0.744 \\
$\lambda_1 $=0.25           & 0.705 & 0.696 & 0.714 & 0.739 & 0.805 & 0.836 & 0.749 \\
$\lambda_1 $=0.5            & 0.702 & 0.695 & 0.718 & 0.740 & 0.812 & 0.836 & 0.751 \\
$\lambda_1 $=1              & 0.704 & 0.696 & \textbf{0.719} & 0.741 & \textbf{0.815} & 0.836 & \textbf{0.752} \\
$\lambda_1 $=2              & 0.704 & 0.698 & \textbf{0.719} & \textbf{0.743} & 0.814 & 0.837 & \textbf{0.752} \\
$\lambda_1 $=4              & 0.703 & 0.699 & 0.718 & \textbf{0.743} & 0.812 & 0.838 & \textbf{0.752} \\
$\lambda_1 $=10             & 0.695 & 0.696 & 0.712 & 0.737 & 0.806 & \textbf{0.840} & 0.748 \\
 \hline
\end{tabular}

\label{tab:lambda1}
\end{table}

\textbf{Block cross-validation.} We explore the effectiveness of employing spatial block cross-validation in determining the optimal value for the pseudo-absence weight $\lambda_2$. We compare this approach with traditional cross-validation methods, including plain cross-validation (randomly splitting presence data) and species-stratified cross-validation, where the proportion of presences is maintained for each species in each subset~\citep{sechidis2011stratification}. During this process, we include the target-group background points associated with the presences in the validation set, along with (optionally) an equivalent number of random background points. The objective is then to identify the $\lambda_2$ value that maximizes the average validation AUC across folds. This is achieved through a grid search for $\lambda_2$, exploring its values within the set $\{0, 0.2, 0.4, 0.6, 0.8, 1\}$.

Results in Table~\ref{tab:crossvalidation} indicate that block cross-validation outperforms other approaches when both random and target-group background points are utilized. This is particularly effective in regions with significant sampling bias, such as AWT, CAN, and SWI. However, superior average performance is achieved by restricting the validation set to include only target-group background points, regardless of the validation method used. Notably, all methods converge on the same selection of $\lambda_2$ values.

\begin{table}[]
\centering
\caption{Mean AUC over the species for different methods employed to construct the validation set used to determine the optimal value of the pseudo-absence weight $\lambda_2$. $\lambda_1$ is fixed at $1$, and $w_s$ is used. The best mean AUC in each column is underlined when both target-group and random background points are included in the validation set. When exclusively using target-group background points in the validation set, all methods converge on the selection of identical $\lambda_2$ values.}

\setlength{\tabcolsep}{0.46\tabcolsep}  
\renewcommand{\arraystretch}{1.2} 
\begin{tabular}{lccccccc}
                                    & AWT   & CAN   & NSW   & NZ    & SA    & SWI   & \textbf{avg}   \\ \hline \hline
\textbf{target-group + random} \\
plain                              & 0.673 & 0.634 & \underline{0.720} & 0.740 & 0.812 & 0.824 & 0.734 \\
species-stratified                 & 0.691 & 0.634 & \underline{0.720} & 0.740 & \underline{0.815} & 0.824 & 0.737 \\
block                              & \underline{0.704} & \underline{0.696} & \underline{0.720} & \underline{0.741} & 0.811 & \underline{0.831} & \underline{0.751} \\
\hline
\textbf{target-group} \\
plain              & 0.704 & 0.714 & 0.719 & 0.741 & 0.815 & 0.838 & 0.755 \\
species-stratified         & 0.704 & 0.714 & 0.719 & 0.741 & 0.815 & 0.838 & 0.755 \\
block              & 0.704 & 0.714 & 0.719 & 0.741 & 0.815 & 0.838 & 0.755 \\
\hline

\end{tabular}

\label{tab:crossvalidation}
\end{table}

\textbf{Additional metrics.} In Tables~\ref{tab:comparison_cor} and ~\ref{tab:comparison_aucprg}, we present the mean correlation and the mean area under the precision-recall gain curve across species, respectively. The results are consistent across the different metrics. Remarkably, employing our full weighted loss with the fine-tuned value of $\lambda_2$ consistently yields the best average performances over the regions.
\begin{table}[]
\centering
\caption{Mean Pearson correlation coefficient over the species. The best mean correlation in each column is highlighted in bold, while the second-best mean correlation is underlined.}

\setlength{\tabcolsep}{0.46\tabcolsep}  
\renewcommand{\arraystretch}{1.2} 
\begin{tabular}{lccccccc}
                                                                          & AWT   & CAN   & NSW   & NZ    & SA    & SWI   & \textbf{avg}   \\ 
    \hline \hline
    \textbf{\cite{cole2023spatial} losses} \\
    SSDL loss, i.e., $\lambda_1 = 1$ and $\lambda_2=0$                    & 0.174 & 0.048 & 0.088 & 0.138 & 0.211 & 0.251 & 0.152 \\
    SLDS loss, i.e., $\lambda_1 = 1$ and $\lambda_2=1$                    & 0.173 & 0.146 & 0.104 & 0.129 & 0.210 & \textbf{0.310} & 0.179 \\
    Full loss, i.e., $\lambda_1 = S/2$ and $\lambda_2=0.5$                & \underline{0.296} & 0.142 & \textbf{0.191} & \underline{0.179} & \underline{0.319} & 0.267 & 0.232 \\
    \hline \hline
    \textbf{Full weighted loss (ours)} \\    
    $\lambda_1=1 $ and $ \lambda_2=0.8$, no $w_s$                         & 0.226 & \textbf{0.178} & 0.119 & 0.144 & 0.238 & \textbf{0.310} & 0.203 \\
    \hline
    $\lambda_1=1 $ and $ \lambda_2=0$, with $w_s$                         & 0.223 & 0.042 & 0.153 & 0.171 & 0.296 & 0.222 & 0.184 \\
    $\lambda_1=1 $ and $ \lambda_2=0.2$, with $w_s$                       & 0.254 & 0.091 & 0.179 & \underline{0.179} & 0.311 & 0.249 & 0.210 \\
    $\lambda_1=1 $ and $ \lambda_2=0.4$, with $w_s$                       & 0.279 & 0.117 & \underline{0.189} & \textbf{0.182} & 0.318 & 0.264 & 0.225 \\
    $\lambda_1=1 $ and $ \lambda_2=0.6$, with $w_s$                       & \underline{0.296} & 0.137 & \textbf{0.191} & \textbf{0.182} & \textbf{0.320} & 0.274 & 0.233 \\
    $\lambda_1=1 $ and $ \lambda_2=0.8$, with $w_s$                       & \textbf{0.300} & 0.158 & 0.187 & \underline{0.179} & 0.318 & 0.280 & \underline{0.237} \\
    $\lambda_1=1 $ and $ \lambda_2=1$, with $w_s$                         & 0.290 & \underline{0.172} & 0.180 & 0.176 & 0.314 & \underline{0.283} & 0.236 \\
    \hline
    $\lambda_1=1 $ and fine-tuned $\lambda_2$, with $w_s$                 & \textbf{0.300} & \underline{0.172} & 0.187 & \underline{0.179} & 0.318 & \underline{0.283} & \textbf{0.240} \\
    \hline
\end{tabular}
\vspace{0.1cm} \\

\label{tab:comparison_cor}
\end{table}
\begin{table}[]
\centering
\caption{Mean area under the precision-recall gain curve (AUPRG) over the species. The best mean AUPRG in each column is highlighted in bold, while the second-best mean AUPRG is underlined.}

\setlength{\tabcolsep}{0.46\tabcolsep}  
\renewcommand{\arraystretch}{1.2} 
\begin{tabular}{lccccccc}
                                                                          & AWT   & CAN   & NSW   & NZ    & SA    & SWI   & \textbf{avg}   \\ 
    \hline \hline
    \textbf{\cite{cole2023spatial} losses} \\
    SSDL loss, i.e., $\lambda_1 = 1$ and $\lambda_2=0$                    & 0.285 & -1.825 & -0.495 & -0.138 & 0.526 & 0.764 & -1.470 \\
    SLDS loss, i.e., $\lambda_1 = 1$ and $\lambda_2=1$                    & 0.198 & -0.505 & -0.687 & -0.546 & 0.511 & 0.843 & -0.031 \\
    Full loss, i.e., $\lambda_1 = S/2$ and $\lambda_2=0.5$                & 0.410 & -0.552 & \textbf{0.203} & \textbf{0.404} & 0.721 & 0.833 & 0.337 \\
    \hline \hline
    \textbf{Full weighted loss (ours)} \\  
    $\lambda_1=1 $ and $ \lambda_2=0.8$, without $w_s$                         & 0.339 & \textbf{-0.036} & -0.300 & -0.451 & 0.602 & \underline{0.846} & 0.167 \\
    \hline
    $\lambda_1=1 $ and $ \lambda_2=0$, with $w_s$                         & 0.291 & -1.267 & -0.107 & 0.320 & 0.690 & 0.759 & 0.114 \\
    $\lambda_1=1 $ and $ \lambda_2=0.2$, with $w_s$                       & 0.340 & -1.057 & 0.123 & 0.355 & 0.710 & 0.800 & 0.212 \\
    $\lambda_1=1 $ and $ \lambda_2=0.4$, with $w_s$                       & 0.385 & -0.782 & 0.186 & 0.374 & 0.722 & 0.816 & 0.284 \\
    $\lambda_1=1 $ and $ \lambda_2=0.6$, with $w_s$                       & 0.427 & -0.629 & \underline{0.202} & 0.384 & \underline{0.728} & 0.829 & 0.324 \\
    $\lambda_1=1 $ and $ \lambda_2=0.8$, with $w_s$                       & \underline{0.456} & -0.430 & 0.166 & \underline{0.388} & \textbf{0.730} & 0.839 & \underline{0.358} \\
    $\lambda_1=1 $ and $ \lambda_2=1$, with $w_s$                         & \textbf{0.461} & \underline{-0.427} & 0.076 & \underline{0.388} & \textbf{0.730} & \textbf{0.847} & 0.346 \\
    \hline
    $\lambda_1=1 $ and fine-tuned $\lambda_2$, with $w_s$                 & \underline{0.456} & \underline{-0.427} & 0.166 & \underline{0.388} & \textbf{0.730} & \textbf{0.847} & \textbf{0.360} \\
    \hline
\end{tabular}
\vspace{0.1cm} \\

\label{tab:comparison_aucprg}
\end{table}

\end{document}